# A 262 TOPS Hyperdimensional Photonic AI Accelerator powered by a Si3N4 microcomb laser

Christos Pappas[1,2,*], Antonios Prapas[1,2], Theodoros Moschos[1,2], Manos Kirtas[1,3], Odysseas Asimopoulos[2,4], Apostolos Tsakyridis[1,2], Miltiadis Moralis-Pegios[1,2], Chris Vagionas[1,2], Nikolaos Passalis[3,5], Cagri Ozdilek[6], Timofey Shpakovsky[6], Alain Yuji Takabayashi[6], John D. Jost[6], Maxim Karpov[6], Anastasios Tefas[1,3] and Nikos Pleros[1,2]

[1]*Department of Informatics, Aristotle University of Thessaloniki, 54124, Thessaloniki, Greece*
[2]*Center for Interdisciplinary Research and Innovation, Balkan Center, 57001, Greece*
[3]*Computational Intelligence and Deep Learning Group, AUTH, 54124, Thessaloniki, Greece*
[4]*Department of Physics, Aristotle University of Thessaloniki, 54124, Thessaloniki, Greece*
[5]*Department of Chemical Engineer, Faculty of Engineer, Aristotle University of Thessaloniki, 54124, Thessaloniki, Greece*
[6]*Enlightra, Rue de Lausanne 64, Renens, 1020, VD, Switzerland*

*\*chripapp@csd.auth.gr*

**Abstract:** The ever-increasing volume of data has necessitated a new computing paradigm, embodied through Artificial Intelligence (AI) and Large Language Models (LLMs). Digital electronic AI computing systems, however, are gradually reaching their physical plateaus, stimulating extensive research towards next-generation AI accelerators. Photonic Neural Networks (PNNs), with their unique ability to capitalize on the interplay of multiple physical dimensions including time, wavelength, and space, have been brought forward with a credible promise for boosting computational power and energy efficiency in AI processors. In this article, we experimentally demonstrate a novel multidimensional arrayed waveguide grating router (AWGR)-based photonic AI accelerator that can execute tensor multiplications at a record-high total computational power of 262 TOPS, offering a ~24× improvement over the existing waveguide-based optical accelerators. It consists of a 16×16 AWGR that exploits the time-, wavelength- and space- division multiplexing (T-W-SDM) for weight and input encoding together with an integrated Si3N4-based frequency comb for multi-wavelength generation. The photonic AI accelerator has been experimentally validated in both Fully-Connected (FC) and Convolutional NN (NNs) models, with the FC and CNN being trained for DDoS attack identification and MNIST classification, respectively. The experimental inference at 32 Gbaud achieved a Cohen's kappa score of 0.867 for DDoS detection and an accuracy of 92.14% for MNIST classification, respectively, closely matching the software performance.

## 1. Introduction

The rapid advancement of artificial intelligence (AI) and deep learning (DL) compute models has signalled the end of Moore's law [1]. Over the last decade, AI has driven computational power demands to double every 5-6 months [2] and pushed daily energy consumption into the hundreds of megawatt-hours [3], highlighting the critical need for a paradigm shift into a novel computing hardware capable of sustaining this massive compute and energy growth. Within this framework, photonic neural networks (PNNs) have been theoretically predicted to hold the credentials for addressing these computational and energy challenges towards enabling Peta-scale compute power and fJ/OP-scale energy efficiency [4]-[6], leveraging their high-bandwidth and low-power characteristics with their constantly growing integration maturity. However, the transfer of theoretical predictions into experimental demonstrations [6]-[31] has revealed a different reality for PNNs, primarily marked by architectural challenges towards safeguarding scalable layouts. Given the footprint handicap of integrated photonic components against their electronic counterparts, PNNs can address scalability only through architectural schemes that can seamlessly support the interplay between time- and space-dimensions together with the traditional optics advantages, i.e. the wavelength-dimension and high-speed operation, which is, however, still absent in state-of-the-art integrated photonic Matrix-Vector-Multiply (MVM) architectures.

More specifically, the demonstrations in [7],[16], exploit photonic meshes and spatial division multiplexing (SDM), executing computations via the use of cascaded Mach-Zehnder interferometer (MZI) nodes. This approach directly correlates physical with computational space, implying a scalability plateau since larger networks would inevitably result in higher insertion losses [32]. On the other hand, extensive research has been conducted on PNN architectures that utilize the wavelength division multiplexing (WDM) technique [15],[26]. The main building block in these architectures is the microring resonator (MRR) bank, which comprises multiple MRRs flanked by two parallel waveguides and being responsible for implementing channel-selective weighting. Despite their impressive performance in a range of applications, the computational power and circuit size of these layouts can scale only by increasing the number of laser sources. At the same time, they necessitate the simultaneous operation and precise control of numerous resonant devices, raising an additional concern regarding their power consumption and scalability perspectives. Photonic MVM circuit size can enjoy unlimited scaling only by exploiting time-division-multiplexing (TDM), with the authors in [12],[14],[20],[27] demonstrating the effective synergy of SDM and TDM by incorporating high-speed input/weight nodes and time integrating receivers for the accumulation operation. Consequently, the size of the PNN is effectively increased without the need to fabricate large photonic circuits, while the employment of the time integrating receiver enables the use of low-power and low-cost analog to digital converters (ADCs) as it relaxes their bandwidth requirements. Yet, these architectures neglect the use of wavelength division multiplexing (WDM) that forms the typical capacity parallelization factor in optics, limiting the ability of PNNs to fully exploit all available degrees of freedom and hence potentially boost their computational power and energy efficiency metrics. The benefits of additionally incorporating the wavelength dimension in time-space division multiplexed setups have been highlighted in more recent demonstrations [23],[28],[29], leading also to computational powers up to 11TOPS [23] that comprise the current record among all state-of-the-art waveguide-based optical processors so far. To the best of our knowledge, only diffractive optics have managed to break the 100 TOPS computational power barrier [33]-[35], either by chiplets built upon diffractive principles [33] or by a hybrid combination of diffractive regions with inference modules [34],[35]. However, the use of diffractive elements requires the a priori encoding of weighting values during the fabrication process and as such inherently constraints photonic AI accelerators to a certain inference task. To this end, the high computational powers enabled by diffraction-based optical processors come at the expense of their flexibility and universality, restricting their deployment only in application-specific inference applications and abandoning the general-purpose character.

In this paper, we experimentally present the first photonic AI accelerator that is powered by a microcomb laser source and is capable of executing Matrix-by-Tensor-Multiply (MbTM) operations at a record-high computational power of 262 TOPS. The proposed architecture consists of a 16×16 AWGR module, broadband intensity modulators for weight and input encoding at 32 Gbaud and a $Si_3N_4$ frequency comb laser source for multi-wavelength generation. This approach extends our recent work on the first AWGR-based PNNs [36] demonstrating a more than 60% increase in total computational power performance and a ~8.5% higher classification accuracy by exploiting 32G clock-rates and microcomb-laser-generated wavelength channels. The topology relies on the cyclic wavelength routing properties of the AWGR and supports simultaneously time-, wavelength- and space- division multiplexing (TWSDM) for weight and input vector encoding, forming a powerful framework for matrix and tensor multiplications. To validate the performance of the proposed tensor accelerator in both FC and CNN layouts, two Deep Learning (DL) models were trained for executing different applications while taking into account the optics-informed DL training framework [4] to adapt to the underlying constraints of the photonic hardware like noise [13],[37], quantization [38],[39] and value non-negativity [40]. The first application was the identification of distributed denial of service (DDoS) attacks in a fully connected NN (FCNN), where the accelerator achieved an experimental Cohen's kappa score of 0.8677 for the 2048 inferenced samples, closely matching the software performance. The second task concerned the classification of handwritten digit images (MNIST) through a convolutional NN (CNN) with the hardware inference revealing an accuracy of 92.14%, showing only 1.55% degradation compared to the accuracy achieved via the software.

## 2. Matrix-by-Tensor Multiplicator Concept and Principle of Operation

The proposed matrix-by-tensor multiplication engine in its generic N×K×S arrangement is illustrated in the conceptual layout of Fig. 1(a). It comprises an $N \times N$ AWGR module with $N$ and $K$ broadband modulators connected to its #N input and #K output ports (K<=N), respectively. A multi-λ stream, generated by multiplexing $N$ laser beams, is split into $N$ channels of equal powers, with the $i$-th channel (i ∈ [1, N]) entering the $i$-th broadband modulator. The modulator is electrically driven by a weight row vector $\overrightarrow{W_i}(t)$, which contains L elements and is optically imprinted on all #N wavelengths, so that the output of each $i$-th modulator consists of an L-symbol long time series. This implies that the supported weight matrices W can have $N$ rows and $L$ columns, with every weight row vector $W_i$ consisting of L discrete symbols that correspond to the number of discrete time slots allocated for its representation. The AWGR collects the outputs of all *i* modulators at its input ports and allows all #N different weight row vectors $\vec{W_1}(t), ..., \vec{W_N}(t)$ to emerge at each AWGR output port by taking advantage of its wavelength cyclic routing properties. Consequently, the whole *N×L* weight matrix W appears at every *j*-th output ($j \in [1, K]$), with every weight row vector carried by a different wavelength within the multi-λ stream that emerges at the same AWGR output port. A splitter with *1:S* ratio at every *j*-th AWGR output port is then employed to broadcast the multi-λ weight matrix into *S spatially separated* copies of equal power level. The resulting signals propagate through a $X_{jl}$ broadband modulator that is placed at every splitter output, with $l \in [1, S]$. The $\vec{X_{jl}}(t)$ electrical input column vectors drive the $X_{jl}$-modulators so that the time instances match the respective *L*-symbol length of the weight vectors. Therefore, the multi-λ signal emerging at the output of the $X_{jl}$-modulator carries the Hadamard product (HP) $\vec{W_i}(t) \circ \vec{X_{jl}}(t)$ between the $\vec{X_{jl}}(t)$ input time vector and all weight time vectors. By demultiplexing the multi-λ signal arising at the output of the $X_{jl}$-modulator into its wavelength-components, one obtains every *L*-symbol long HP between a single weight row vector and the input vector. Accumulation is then performed by employing an optical or optoelectronic integrator at every demultiplexer output, which integrates over a *L*-symbol time duration to provide the dot-product, $Y_{i,jl}$, between the vectors $\vec{W_i}(t)$ and $\vec{X_{jl}}(t)$, as proposed in [41]. In this way, a different MbMM, between the weight matrix W and the respective input matrix formed by the *S* different $\vec{X_{jl}}(t)$ column vectors, is calculated for each *j*-th AWGR output, as shown in Fig. 1(b). By exploiting the spatial dimension and equipping all *K* AWGR output ports with a similar 1:S split-and-modulation MbMM stage, the proposed architecture can successfully execute a number of *K* different MbMM operations effectively turning the setup into a MbTM layout where a *N×L* weight matrix gets multiplied by a *L×S×K* input signal tensor. In this way, assuming a typical case where *K=N* for the AWGR input and output ports, the proposed layout supports a total number of $N^2 \cdot S$ computations. By assuming additionally an operational baud-rate *B* Gbaud at every modulator and a splitting ratio *S* that equals the number of ports *N*, then the number of computations scales as $O(N^3)$ although the circuit complexity scales as $O(N^2)$, resulting to a total computational power of $N^3 \cdot B$ GMAC/sec or, equivalently, $2 \cdot N^3 \cdot B$ GOPS. Figure 1(c) pictorially represents the envisioned AWGR-based architecture, showcasing all key building blocks comprising the proposed topology.

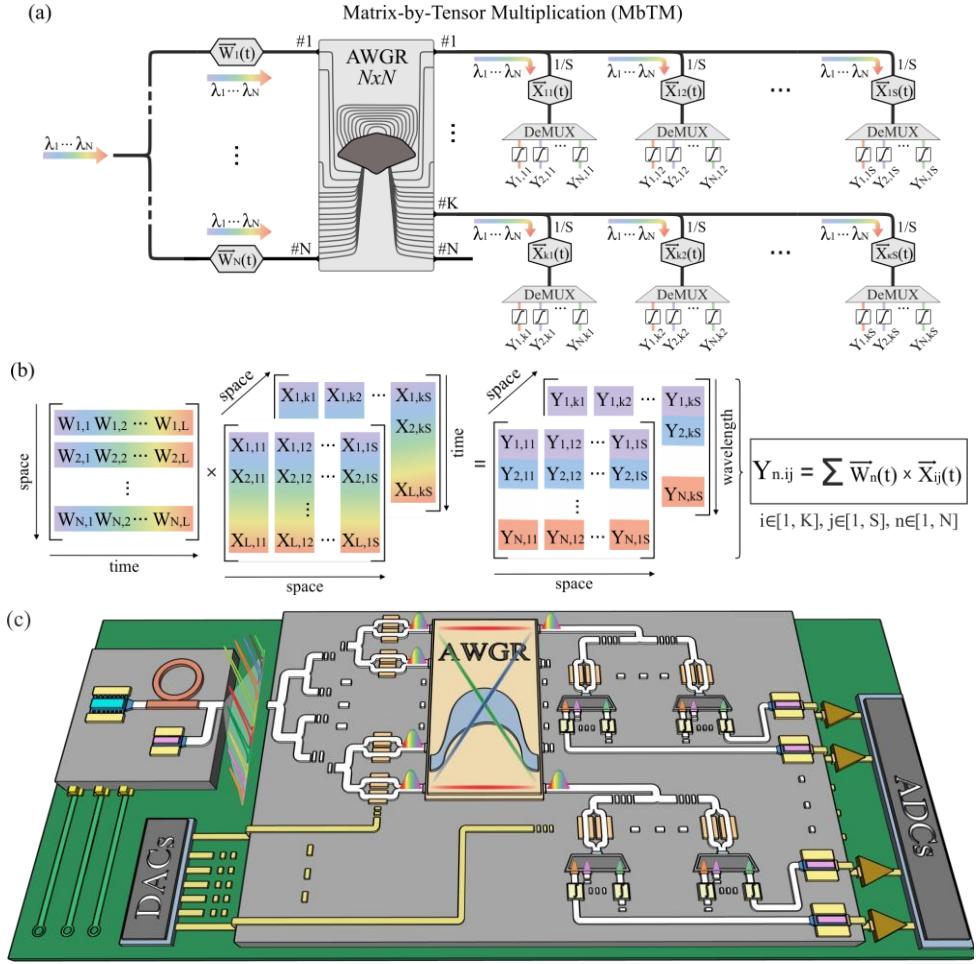

Figure 1: (a) Conceptual layout of the AWGR-based Matrix-by-Tensor-Multiplication, (b) the corresponding W-matrix and X-tensor with their respective results $Y_{ij}$, and (c) the envisioned integrated AWGR-based accelerator.

## 3. Experimental Implementation of the AWGR-based Accelerator

The experimental implementation of the AWGR-based photonic MbTM engine is illustrated in Fig. 2(a). A silicon nitride ($Si_3N_4$)-based frequency comb source (Enlightra's SLC) was employed as a continuous wave (CW) generator, with the presented block diagram revealing the basic building blocks: i) a CW pump laser, ii) a high-quality $Si_3N_4$ micro-resonator, and iii) a controlling unit for the electronics. Light coupled into the ring cavity builds in intensity with each roundtrip pass until the circulating optical power exceeds the material's nonlinear threshold to generate a wide range of frequencies, i.e., a frequency comb, with a free spectral range (FSR) of 100 GHz. The combination of high Kerr nonlinearity in silicon nitride, group velocity dispersion and high-Q factor around 1 million of the microresonator enables the generation of a highly coherent optical frequency comb through a cascaded four-wave mixing processes. The generated optical spectrum is centered at 1550 nm and features a 3-dB bandwidth of >2 THz with an OSNR reaching 50 dB. An integrated feedback loop with built-in photodiodes (PD on the figure) and electronics ensures long-term stabilization of the system, with >2000 hours of non-stop operational stability having confirmed experimentally. An external graphical user interface allows control of the internal components and parameters of the frequency comb. Enlightra's SLC was emitting a total optical power of 1mW, for all the produced wavelengths, with an erbium doped fiber amplifier (EDFA) connected at the output of the device boosting the optical power. The CW generation stage was followed by a demultiplexing-multiplexing stage allowing for a per channel adjustment of the selected

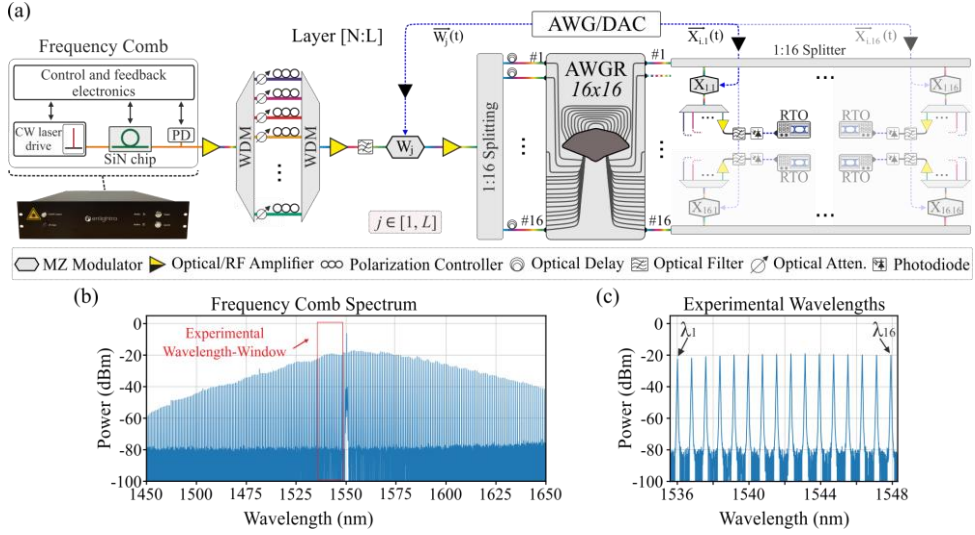

Figure 2: (a) AWGR-based experimental testbed for MbTM products verification, (b) frequency comb output spectrum, with the red rectangle denoting the experimental wavelength-window, and (c) zoomed-in spectrum of the experimental wavelengths.

wavelengths' power levels ensuring the generation of 16 power-equalized channels with a channel spacing of 0.8 nm within the spectral window of 1535.9 - 1547.8 nm, with a spacing error of ±0.01 nm. The polarization state of every channel was also on a per channel basis adjusted to allow for efficient amplitude modulation of all 16 wavelengths in the subsequent modulation stage. A second EDFA was placed after the multiplexer for loss compensation, with a bandpass filter (BPF) of 12 nm optical bandwidth removing the amplified spontaneous emission noise. The filtered multi-λ stream with total average power of 28.1 mW, or ~1.8 mW per wavelength-channel, was then injected into an indium phosphide (InP) Mach-Zehnder modulator (MZM) with 35 GHz electro-optical (EO) bandwidth (Fraunhofer HHI InP MZM 64 Gbaud), that was driven by an arbitrary waveform generator (AWG – Keysight M8194A) to generate the $\overrightarrow{W_j}(t)$ row vector of the W-matrix. The AWG generated the electrical signal with a power of 350 mV, which was further amplified via an RF amplifier (SHF M804B), reaching ~3.5 V as necessitated from the InP MZM. In this way, an $L$-value weight row vector $\overrightarrow{W_j}(t)$ was imprinted as a time series onto all 16 wavelength-channels. After optically amplifying the W-modulator output to compensate for losses in the subsequent components, the signal was split into 16 identical spatial components via a 1:16 splitting stage, with the 16 WDM signals getting decorrelated via the employment of different fiber-length delays at every of the 16 paths prior entering the respective AWGR input. The employed AWGR (Semicon SAWG-G-100G-32-32-C) is commercially available and consists of 32 input and 32 output ports with a 3 dB bandwidth of 0.4 nm. For this experiment though, we only exploited the 16 input and 16 output ports acting in this way as a 16×16 AWGR. As such, the $\overrightarrow{W_j}(t)$ row vector entering the AWGR via a certain input port will emerge at every of the 16 AWGR output ports but at a different wavelength at each port due to the cyclic wavelength routing properties. This means that every AWGR output port will carry 16 different wavelengths with every wavelength stemming from a different AWGR input and carrying a time-delayed copy of the $\overrightarrow{W_j}(t)$ row vector. In this way, only one wavelength can be considered at a certain AWGR output to carry the targeted $\overrightarrow{W_j}(t)$ row vector with the rest 15 wavelength-channels exiting through the same AWGR output aggregating the channel under evaluation. The multi-λ modulated stream derived at the $k$-th AWGR output was further split by a 1:16 splitter creating 16 identical copies at respective paths. Each $i$-th splitter output port was fed to a LiNbO₃ $X_{ki}$-input MZM (iXBlue MX-LN-40) with 40 GHz EO bandwidth, designated to modulate the corresponding $\overrightarrow{X_{ki}}(t)$ input vector. The LiNbO₃

MZM was driven again by the AWG module to produce the optical $X_{ki}$ input vector time series, and as before, a RF amplifier (SHF L806A) was employed to amplify the 200-mV electrical signal generated via the AWG. By sequentially connecting the $X_{ki}$-input MZM to the 16 splitter outputs of the *k*-th AWGR output port, we successfully acquire the MbMM product. After repeating the above procedure for the remaining AWGR ports, by successively connecting the 1:16 split-and-modulate AWGR output stage to all AWGR outputs, the required MbTM products are obtained. To evaluate the different Hadamard products carried by every wavelength, the multi-λ signal was demultiplexed into its 16-wavelength constituents at the output of every $X_{ki}$-MZM. Every channel was then amplified in an EDFA followed by a BPF with 0.55 nm bandwidth prior entering a 70 GHz photodiode (PD – Finisar XPDV3120) and being recorded by a 256 GSa real-time oscilloscope (RTO – Keysight UXR0704AP). A software-based filter was implemented via the RTO and applied to the captured signal, with a manually adjusted 3 dB bandwidth of 20 or 32 GHz in order to mitigate the excess noise bandwidth of the PD. The integration and non-linear activation function (AF) were performed off-line via a respective software routine.

## 4. Experimental Validation in DL applications

The experimental validation of the proposed MbTM architecture was carried out in two different AI applications that exploit two different DL models and respective NN configurations. The scope of the two applications was: (i) cybersecurity, through the identification of DDoS attacks in Data Centers (DCs) via the analysis of data packet traffic and the classification between malicious and benign packets, and (ii) classification of handwritten digits, using the MNIST dataset. The training procedure for the DDoS identification and MNIST classification tasks started with the software-based implementation of the task-specific NN models, i.e., an FC for the DDoS and a CNN for the MNIST, using the PyTorch framework. In parallel, a simulation of the photonic architecture under testing extracted the mathematical definition of the hardware-based constraints so they could be included in the NN training process. The described pipeline is elaborated in [38].

The first step was to apply a linear dimensionality reduction to the DDoS dataset using Principal Component Analysis (PCA) to reduce the input vector size of the telemetry data to 6. The extraction process of the characteristics for the DDoS dataset is further detailed in [42]. After each layer, the digital sigmoid AF provided the non-linearity of the system. The NN was trained with the Adam optimizer [43] for 100 epochs, applying hard constraints to weights, forcing them on non-negative values, within the range of [0, 1]. These weight constraints were incorporated during training by introducing a penalty term to the loss function, formulated as:

$$L(\hat{y}, y; w) = L_c(\hat{y}, y) + \alpha \sum_{k=0}^{K} \sum_{i=0}^{N_k} \sum_{j=0}^{M_k} \frac{\sqrt{\max\left\{-w_{ij}^{(k)}, 0, w_{ij}^{(k)} - 1\right\}^2}}{N_k M_k}, Eq. (1)$$

where $L_c$ denotes the cross-entropy loss and *α* is the weighting factor for the penalty term, which by default is set to $\alpha = 10$. The $N_k$ and $M_k$ define the fan-in and fan-out of the *k*-th layer, respectively. Additionally, the hard constraints for the clipping of the weights in the range of [0, 1] were applied after the 10-th epoch, following the optimization step.

The implementation of the CNN for MNSIT classification employed three convolutional software-based layers, two convolutional hardware-based layers and one fully connected hardware-based layer. The software-based convolutional layers introduced a backbone for feature extraction and consisted of three layers. The first two (Conv 1, 2) employed a 3×3 kernel with 3 channels each and applied the digital ReLU AF as the non-linearity. The third (Conv 3), employed a 3×3 kernel with 4 channels, and with stride and dilation equal to 2, with the non-linearity implemented by the digital sigmoid AF. The backbone's output, a 4-channel 11×11 feature map, fed into the cascaded layers, which were validated through the optical hardware. The first optical convolutional layer (Conv 4) applied a 4×4 kernel with 8 channels and a stride of 4. The extracted feature map continued to the second optical convolutional layer (Conv 5) that applied a 2×2 kernel with 32 channels. The output of the last photonic convolutional layer (Conv 5) was flattened to a column-vector, consisted of 32 values ready to be processed by the optical classification/output layer (FC). The non-linearity selected for the hardware-based layers, was experimentally extracted by a programmable opto-electro-optical system with the operational principles detailed in [44].The CNN was trained in an end-to-end manner for 75 epochs, applying different optimization approaches to the software-backbone layers and to the hardware layer. Specifically, the backbone optimization employed the Adam optimizer with a learning rate equal to 0.001 and included a learning rate scheduler to reduce its value by half, every 25 epochs. The optical layers were restricted to 3 bit-precision and employed a normalized quantization-aware training approach [39]. Optimization of the optical layers utilized the multiplicative Adam optimizer [40] to ensure that the non-negativity applied during weight-initialization remained during the training process. In this way, the hard constraints on the optical layers' weights were mitigated using only the upper-bound penalty term of the applied loss function of Eq. 1, with α=1.

Figure 3(a) illustrates an instance of telemetry data of the data traffic within a DC, which were used in the DDoS identification task. The telemetry data were categorized in six classes, corresponding to the six input vectors of the first FC NN layer, as shown in Fig. 3(b), followed by an output FC layer and two outputs. Each of the neurons of layer 1 (L1) and layer 2 (L2) is described by the layout of Fig. 3(c). Prior to evaluation, a preliminary pre-emphasis procedure (discussed in the supplementary material) was employed to compensate for i) the noise originating from the limited frequency response of the deployed modulators and ii) the non-linearities within the electro-optic system. In order to allow the processor architecture to utilize all its weighting and input signal modulators for useful computations without sacrificing any wavelength or modulator resources for negative number representation [45], the DDoS classification NN was trained to allow only for non-negative values through the use of optics-informed DL models [4] (more details in the Methods section). Figure 3 (d) illustrates the software-obtained together with the experimental time traces of the Hadamard products $\vec{W_i}(t) \circ \vec{X_{jl}}(t)$ obtained at the first splitter output connected to the first AWGR-output, at wavelength $\lambda_1$ at two different symbol rates, i.e. 20Gbaud and 32Gbaud, respectively. As can be observed, the experimental traces are closely following the respective software-obtained counterparts. The noise distribution both at 20 and 32 Gbaud are also shown in Fig. 3(d), revealing a standard deviation (STD) of only 0.05 for 20Gbaud that increases to 0.095 when 32Gbaud waveforms are used. Similar results were obtained for all wavelengths across all 16 outputs of the split-and-modulate stage at every AWGR output port. The DDoS attack recognition used a set of 2048 input samples consisting of 81.45% (1668) benign and 18.55% (380) malicious packets, which resulted in a highly imbalanced dataset. The Cohen's kappa-score [27],[46] metric accounts for the imbalance of classes and therefore provided a more accurate representation in the final validation. The confusion matrices resulting from the software-based inference and the hardware when operating at 20 Gbaud and then at 32 Gbaud are presented in Fig. 3(e), along with their respective kappa-scores. The scores obtained with this hardware, i.e., 0.8718 and 0.8677 for the 20- and 32-Gbaud cases, respectively, indicate excellent performance with minor degradation compared to the software case.

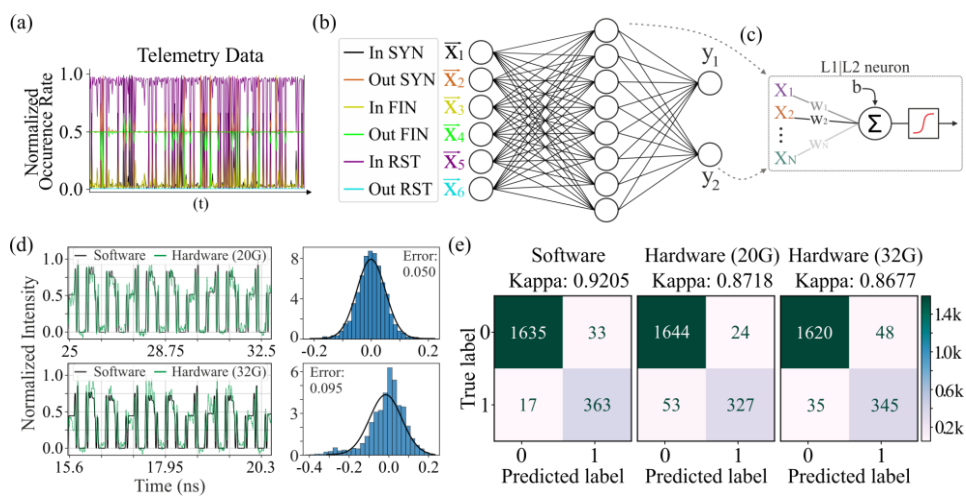

Figure 3: (a) Telemetry data of the DC traffic, (b) fully connected NN topology, (c) L1 and L2 neuron breakdown, (d) time traces for the software experimental traces at 20- and 32Gbaud, with their respective errors, and (e) confusion matrices obtained from the inference performed on software, hardware 20- and 32-Gbaud, along with the respective achieved kappa scores.

To evaluate the performance of the photonic AI accelerator also in CNNs, a second experimental validation process was carried out for the classification of handwritten digits, through a hybrid software/hardware NN. More specifically, the NN topology comprised 5 convolutional layers and a fully connected layer for the final classification. Among these layers, the first 3 convolutional layers were executed in software and were responsible for the initial dimensionality reduction, while the remaining two convolutional layers and the last fully-connected layer were implemented over the photonic hardware. The NN topology and the respective size of each layer, are illustrated in Fig. 4(a). Again, an optics-informed DL training scheme was employed to allow for the use of strictly non-negative values and take into account the different value, quantization and noise constraints of the photonic hardware (more details in the Methods section). Figure 4(b) depicts an indicative time trace at the output of the CNN layer 5 (Conv 5) and the FC layer for the software and hardware multiplications, after the software correlation, for a time-window of 4.7 ns or 150 symbols, at 32 Gbaud. The experimental traces closely follow those acquired from the software, with similar results being obtained for all different wavelengths across all possible PNN outputs. The noise distribution of all output waveforms is shown in the same figure and reveals a noise STD of only 0.053 and 0.12 for the Conv 5 and FC layer, respectively. This NN task classified a total of 256 images using software and hardware inference with relatively balanced subsets for each class. The confusion matrices of the software- and hardware-based classifications are depicted in Fig. 4(c). The software-based inference for the selected 256 samples reached 94.53% accuracy, and the hardware-based inference, at 32 Gbaud operation achieved 92.14% accuracy. This difference corresponds to a misclassification of only 4 samples for the hardware inference. The kappa-score values were also calculated to account for the slightly imbalanced classes in the dataset, revealing software- and hardware-based values are 0.9389 and 0.9214, respectively.

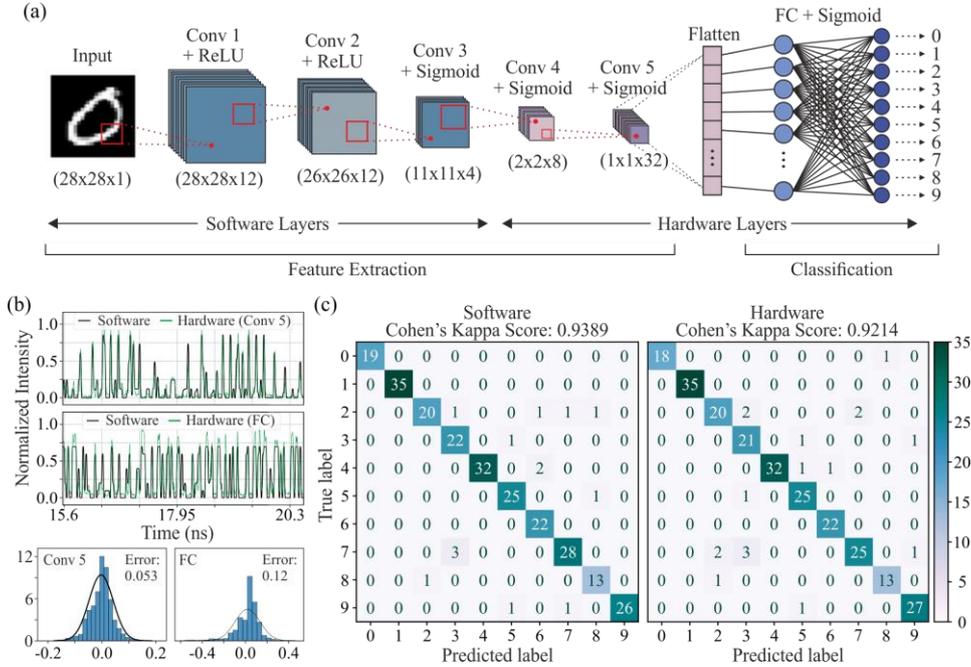

Figure 4: (a) The CNN network used for MNIST classification, with the software and hardware layers being denoted, (b) software and hardware traces at 32 Gbaud for Conv 5 and FC layers, together with the respective errors, and (c) confusion matrices for software and hardware inferences, with their kappa scores.

## 5. Discussion

The proposed AWGR-based multidimensional tensor-multiplication demonstrator, consisted of a 16×16 AWGR module, a frequency comb laser providing the 16 carrier wavelengths and high-speed MZMs. By driving the weight and input signal modulators at 20 Gbaud, the total computational power reaches the value of 163.84 TOPS, which increases to the record-high value of 262 TOPS when increasing the symbol rate at 32 Gbaud, exhibiting a ~60% increase in computational power when compared to our previous work [36] or ~2200% increase when compared to the next, in the exhibited TOPS, PNN architecture [23]. Figure 5 juxtapositions the recent PNN demonstrators in terms of the achieved compute performance in TOPS, where a projection of a future implementation of a 32×32 AWGR-accelerator is included. The demonstrated version of the TSWDM AWGR-based hardware prototype employed fiber-based components but can potentially be transferred to a chip-scale integrated version, taking into account the current capabilities of silicon photonic integration technology. This can certainly lead to improved performance both with respect to the total computational power and to the energy efficiency metrics. Fig. 5 depicts the scalability perspectives of the architecture and plots the two key metrics of computing engines, i.e., computational power (CP) and energy efficiency (EE) versus the AWGR size, considering state-of-the-art integrated photonic components.

The analysis assumes the use of an integrated $N \times N$ AWGR and operation of the PNN at three different data rates: 20 and 32 Gbaud, as presented in this work, and a future target of 50 Gbaud that has been already shown to successfully support AI applications at chip-scale 27. Quantification of the respective EE requires a detailed breakdown of the power consumption (PC) of all its active components, incorporating also the electronic digital-to-analog conversion (DAC), integrator and analog-to-digital conversion (ADC) circuits. The assumptions about speed and energy consumption of a chip-scale PNN version are provided in detail in the Supplementary Material. The MbTM throughput scales with $O(N^3)$ and the achieved operations can be calculated as $2 \cdot N^3 \cdot B$ GOPS, for K=N. As such, the total computational power increases with the cube of the AWGR dimensions, while energy efficiency improves since the total power consumption scales with the total number of active components employed, which in turn scale with $O(N^2)$. The computational power and energy efficiency improve also with increasing data-rate, indicating that a silicon photonic version of the 16×16 AWGR-based PNN layout at 32 Gbaud allows for a total compute power of 262 TOPS with an energy efficiency of 273 fJ/OP. Assuming a 32×32 AWGR 47 silicon-based PNN operating at 50Gbaud symbol rates could then in principle enable a total computational power up to $2 \cdot 32^3 \cdot 50 \cdot 10^9 = 3.276$ POPS, within a power consumption envelope of 309.5 W, suggesting an energy efficiency of ~94 fJ/OP.

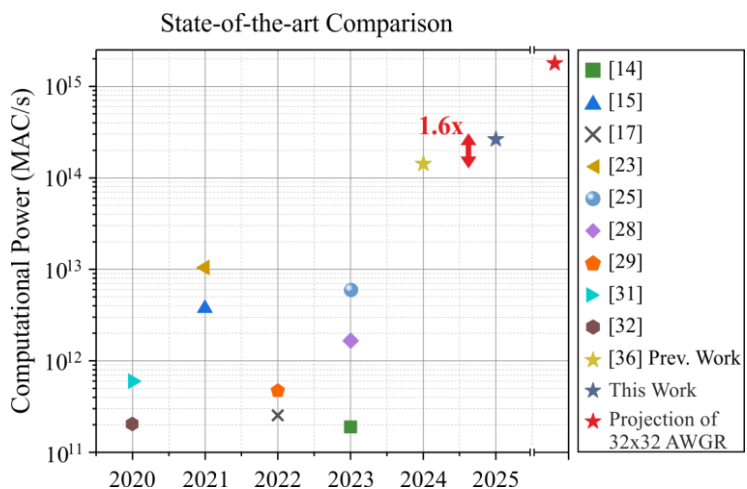

Figure 5: Comparison of state-of-the-art PNN topologies.

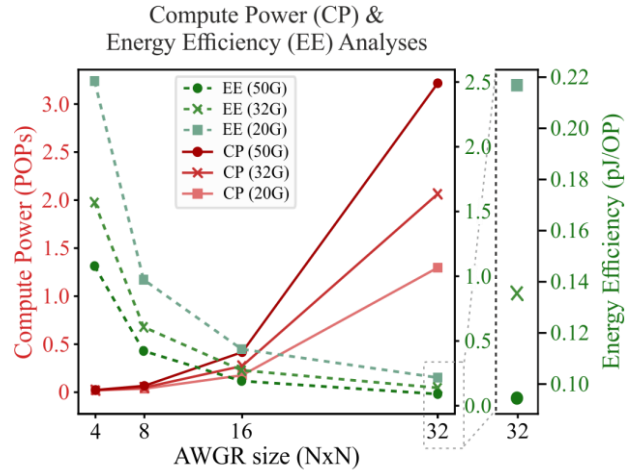

Figure 6: Scalability analysis of compute power and energy efficiency versus AWGR size.

## 6. Conclusions

We have experimentally validated an MbTM architecture that comprises a comb laser source, a 16×16 AWGR and broadband MZMs driven up to 32 Gbaud data-rates, providing a record-high computational power of 262 TOPs. Two different DL models for respective applications were trained and utilized for the experimental validation of the proposed MbTM layout in the AI domain: a FC NN for DDoS identification and a Deep NN formed by multiple convolutional layers and a last FC layer for MNIST classification. Successful classification between benign and malicious traffic in the DDoS attack identification task was experimentally observed, with an experimentally obtained Cohen's kappa-score of 0.8677 over 2048 inference samples with only a 0.05 degradation compared with software. The classification of digits within the MNIST dataset achieved an experimental accuracy of 92.14% over the hardware inference of 256 samples, which closely matched the software performance of 93.89%. Finally, we discussed the integration perspectives of the proposed architecture and highlight this as a promising roadmap towards additional improvements in computational power and energy efficiency. The projected AWGR-based architecture, with upgraded connectivity to 32×32 and a compute rate of 50 Gbaud, is expected to allow for 3.276 POPS of computational power consuming sub-100 fJ/OP, which constitutes a ~298x increase in computational power compared to state-of-the-art photonic accelerators.

## Data Availability

The data that support the findings of this study are available from the corresponding author upon request.

## Acknowledgements

Enlightra would like to thank Charlotte Bost for her work during PIC linear characterization and Lou Kanger for nonlinear characterization. Enlightra would also like to thank Alexey Feofanov for his assistance in the development of the Enlightra SLC. The work was in part funded by the EU-project Gatepost (101120938) and ALLEGRO (101092766).

## Disclosure

The authors declare no conflicts of interest but disclose in the interest of transparency that M.K and J.D.J are co-founders of Enlightra.

## Author Contributions

C.P., T.M., M.M.P., A.T. and N.P. conceived the experiment. C.P., A.P., and T.M. deployed the experimental setup, performed the experiment and processed the experimental results. M.K., O.A., N.P. and A.T. performed the training of the neural network models. C.O. manufactured the device and packaged the photonic integrated circuit. T.S. simulated the design of the photonic integrated circuit and developed the software for the device. A.Y.T., J.D.J. and M.K. supervised the work. C.P., M.K., A.T. and N.P. wrote the manuscript. All authors discussed the results.

# A 262 TOPS Hyperdimensional Photonic AI Accelerator powered by a Si3N4 microcomb laser: Supplemental Document

Christos Pappas, Antonios Prapas, Theodoros Moschos, Manos Kirtas, Odysseas Asimopoulos, Apostolos Tsakyridis, Miltiadis Moralis-Pegios, Chris Vagionas, Nikolaos Passalis, Cagri Ozdilek, Timofey Shpakovsky, Alain Yuji Takabayashi, John D. Jost, Maxim Karpov, Anastasios Tefas and Nikos Pleros

*chripapp@csd.auth.gr

**S1.     Pre-emphasis process towards linearizing the photonic link**

Following the setup procedure of the experimental testbed, and prior executing the multiplication for the NN models, we carried out a preliminary testing for evaluating the performance of the multiplication operations. Initially, each Mach-Zehnder modulator (MZM) of the experimental setup (described in the section: *Experimental Validation through DL applications* of the main manuscript) was driven by a PAM4 electrical signal, capturing their response in a real time oscilloscope (RTO). The initial performance indicated a mismatch between the expected and the received values, as the 4 ideal levels were mixed. This is due to two factors: (i) the noise originating from the limited frequency response of each channel-pair [RF driver - MZM], and (ii) the non-linear response introduced by the RF driver, MZM and PD combination. It is worth noting that the frequency response will change based on the operational data-rate, as opposed to the non-linear response which only accounts for the operational conditions, i.e., RF driver input power and MZM's bias point. To mitigate the effects of both impairments, we investigated pre-emphasizing the signal using offline software at the transmitter site, which is a common digital signal processing procedure. As already mentioned, the nominal data-rate can affect the frequency response. Hence, a standard pre-emphasis of the frequency response of each [RF driver - MZM] pair was incorporated based on the operating speeds of 20 and 32

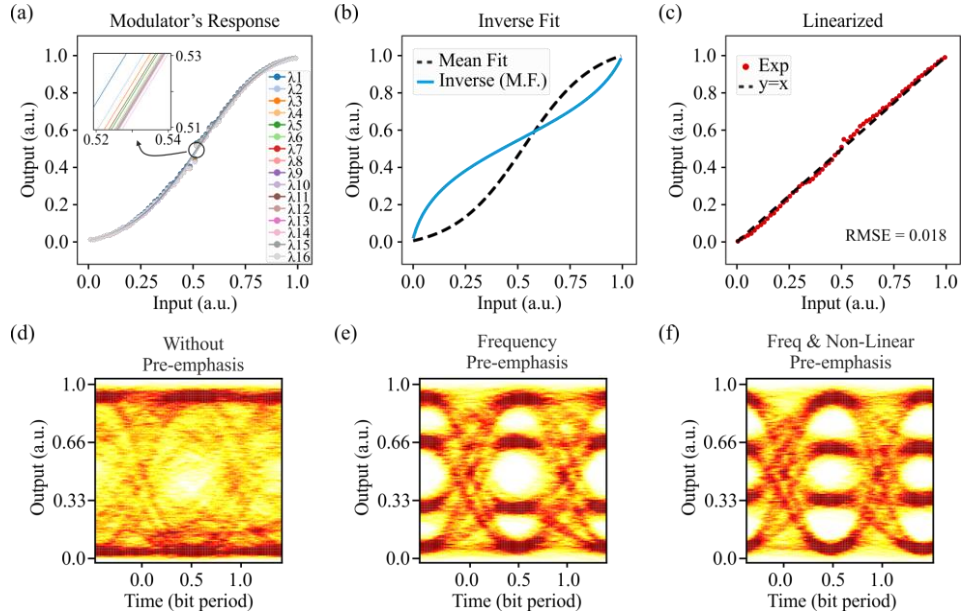

Supplementary Figure 1: Pre-emphasis procedure for frequency and non-linear compensation, (a) non-linear response of the MZM for the 16-wavelength channels with the zoomed-in inset showing the deviation of each wavelength, (b) mean fit of the 16-wavelength response with the respective inverse function applied to the transmitter, (c) linearized response after the non-linear compensation, (d) eye-diagram of a PAM4 signal without applying any pre-emphasis, (e) eye-diagram of the same PAM4 signal by applying the channel-frequency pre-emphasis, and (f) eye-diagram of the same PAM4 signal by applying channel-frequency pre-emphasis and non-linear compensation.

Gbaud, respectively. This pre-emphasis step involves applying a 7-tap digital linear feed-forward equalizer at each MZM output, with the produced taps being enforced to the original signal at the transmitter.

For the non-linear compensation that followed the frequency pre-emphasis, a custom multi-level signal was created that facilitated the procedure of extracting the transfer function (TF) – response of the electro-optic link. The custom electrical signal was designed as a linearly increasing and decreasing sequence of pulses oscillating over a DC value (zero) and was generated as: $Sequence = [0, +n, -n], n \in [1, 32]$. This signal enabled a 64-discrete-level resolution for the TF extraction. The non-linear components we accounted for were the non-linearities introduced by the RF driver, MZM and PD. Although the MZMs are broadband devices, the TF extraction considered the response of each one of the 16 wavelengths exploited experimentally, with the resulting transfer characteristic curves being illustrated in Sup. Fig. 1(a). The 16 wavelength-curves were separately captured for the same input power and bias conditions for the MZM, and were normalized in the range (0, 1). The graph reveals a small deviation between the mid-value points of the curves, as shown in the inset of the figure. For the 16 curves we calculated the mean fit, as presented in Sup. Fig. 1(b). Based on this, we extracted the inverse function and applied it to the ideal signal at the transmitter. The linearized response of the MZM is depicted in Sup. Fig. 1(c), where the experimentally derived points, are fit and closely aligned with the linear function exhibiting an RMSE of only 0.018. Finally, Sup. Figs. 1(d)-(f) illustrate the eye diagrams for the PAM4 signal when the MZM is driven at 20 Gbaud. Fig. 1(d) shows the eye diagram when neither pre-emphasis nor non-linear compensation are applied to the transmitter. Fig. 1(e) illustrates the PAM4 eye diagram when only frequency pre-emphasis applied, while Fig. 1(f) depicts the PAM4 eye diagram, when both frequency pre-emphasis and non-linear compensation are applied, clearly showcasing significant improvement over the original signal. The same procedure was also followed for the case of 32 Gbaud.

**S2.    Multi-wavelength operation validation**

To validate the multi-wavelength operation across the experimental testbed, we obtained the optical spectra at four different points. These points are detailed within the small-scale representation of the experimental setup shown in Sup. Fig. 2(a), labeled as points A-D. These points were aptly selected to follow the basic experimental stages, i.e., after the wavelength (de)-multiplexing, W- and X-modulators and AWGR output. Supplementary Figure 2(b) illustrates the four different spectra, when modulating with 32 Gbaud data-rate. As it can be observed, the power levels among the 16 wavelengths are equalized at each modulation stage, in order to ensure consistent quality performance among the wavelength-carriers of the NN data. The multiplexer output spectrum at point A, shows that the wavelengths near 1536 nm have lower peak powers. This was engineered based on the gain curve of the subsequent erbium doped fiber amplifier (EDFA). The W-modulator output (point B) illustrates almost uniform peak powers across all available wavelengths, validating its broadband capabilities. The peak power distortion between the wavelengths at AWGR output spectrum (point C) owes to the unbalanced losses originating from the I/O ports of the AWGR module. To ensure a peak power-equalization at both the X-modulator input and output (point D), a wave-shaper was placed at the output of the AWGR in order to customize the attenuation per wavelength-channel. It is worth noting that the data rate of 32 Gbaud, was the maximum achievable based on the experimental devices we employed. As shown in Sup. Fig. 2(b) and spectrum-graph of point C, the wavelengths generated from the frequency comb laser were misaligned with the AWGR-channels. Consequently, an increase of the data-rate would result in information-loss due to the grid mismatch of the frequency comb and AWGR-channel. The frequency comb was already thermally tuned to the maximum achievable detuning (wavelength+$\Delta\lambda$), while the heaters of the AWGR module could not be adjusted due to a faulty control circuit.

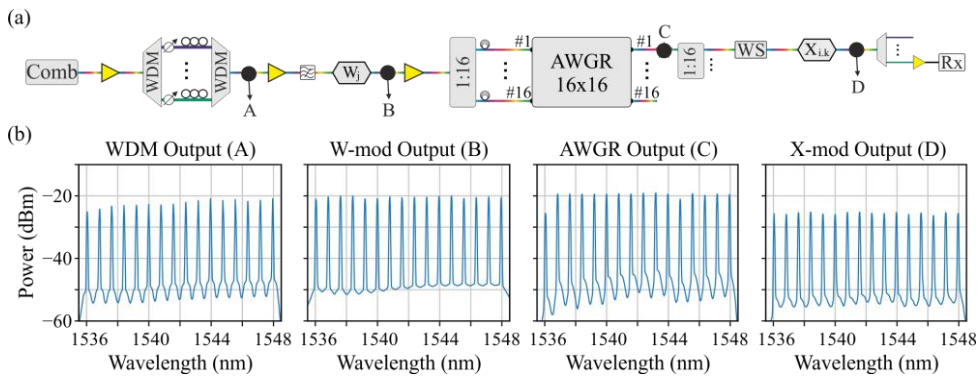

Supplementary Figure 2: (a) Small-scale experimental setup (detailed in main manuscript) denoting 4 different points (A-D) for spectrum capturing, and (b) Spectra at different points of the setup for the multiplexer output (A), W-mod output (B), AWGR output (C) and X-mod output (D).

**S3.    Power consumption calculations and breakdown**

A SiPho/InP-based integrated version of the proposed $N$-channel AWGR architecture, will still follow the structure of the layout presented in Figure 1(a) of the Main Manuscript. This means that the photonic integrated circuit (PIC) would require a multi-wavelength laser source, two splitting stages (a 1: $N$ splitter prior each modulation stage), high speed modulation nodes and the readout circuitry, i.e., transimpedance amplifiers (TIAs), integrators and analog-to-digital conversion (ADC) channels. The most appealing multi-wavelength laser source, in terms of footprint and energy efficiency, is the frequency comb laser, as also incorporated in this experimental demonstrator. Though, the current frequency comb implementation, which we considered for the analysis, is coming with a limitation in terms of the per wavelength-channel average power (~0.01mW) and thus we also need to account for amplification stages within the PIC. To determine the number of amplifiers, the required gain and their placement within the architecture, we first need to breakdown the insertion loss (IL) of the $N$-channel AWGR topology. The $IL_{Total}$ can be calculated as: $IL_{Total} = 2 \times IL_{MZM} + IL_{AWGR} + IL_{DeMUX} + 2 \times IL_{SPLIT(1:N)}$. The only variable term of the equation that can affect the required number of amplifiers and their respective gain, is the term $IL_{SPLIT(1:N)}$, which is based on the variable $N$ of the $N$-channel AWGR module. We considered semiconductor optical amplifiers (SOAs) as the amplification medium, with their total count per $N$-channel case with their respective gain, detailed within Supplementary Table 1. For every $N$-channel case, an SOA of gain 15 dB or 20 dB is assumed to be directly connected at the output of the frequency comb. For the 4- and 8- channel cases we assumed $N$ SOAs connected to the $N$-outputs of the AWGR, while for

**Supplementary Table 1: Power consumption and total count of the SiPho-based components considered for the N×N AWGR architecture.**

| Component | Reference | Consumption (mW) | Component Count |
|---|---|---|---|
| Frequency Comb | | 1,000[*] | 1 |
| DAC | 1 | 144 / 168[†] | $N + N^2$ |
| RF Amplifier | 2 | 100 | $N + N^2$ |
| SOA[†] | 3 | 42 / 84[‡] | **4×4 AWGR:** $N^{(15dB)} + 1^{(15dB)}$ |
| | | | **8×8 AWGR:** $N^{(15dB)} + 1^{(20dB)}$ |
| | | | **16×16 AWGR:** $2 \times N^{(15dB)} + 1^{(15dB)}$ |
| | | | **32×32 AWGR:** $2 \times N^{(15dB)} + 1^{(20dB)}$ |
| TIA | 4 | 9/$L$[⸸] | $N^3$ |
| Integrator | 5 | 0.44[⸸] | $N^3$ |
| ADC | 6 | 0.56[⸸] | $N^3$ |

[*]Consumption is based on the targeted value of the future optimized Enlightra SLC frequency comb.

[†]Consumption is based on the respective data-rate and is calculated as 144 mW (20/32G) or 168 mW (50G).

[†]Different SOA-gain is accounted based on the AWGR's I/O ($N \times N$) at the component count.

[‡]Consumption is based on the respective gain and calculated as 42 mW [15 dB] or 84 mW [20 dB].

[⸸]The readout circuitry needs to operate in a sub-GHz speed, as it will be active only once every $L$ time-steps.

the 16- and 32-channel cases, we considered *N* SOAs connected prior the first modulation stage to compensate for the 12- or 15-dB of IL introduced by the 1:16 or 1:32 splitter, respectively, and *N* SOAs placed at the *N* outputs of the AWGR to compensate for the IL of the subsequent 1:16 or 1:32 splitter. Supplementary Table 1 also contains the device count for the rest components, i.e., DAC, RF amplifier, TIA, integrator and ADC, along with their corresponding power consumption values. Based on the corresponding consumption metrics, the power consumption of the demonstrated 16×16 AWGR-based MbTM accelerator operating at 32 Gbaud data rate, can be quantified as 71.59 W. For the achieved computational power of 262 TOPS this power consumption translates to a respective energy efficiency of ~273 fJ/OP. Finally, the targeted MbTM accelerator will employ a 32×32 AWGR and use modulators with compute rate of 50 Gbaud. In this case, the power consumption adds up to 309.5 W and for the achieved computational power of 3.276 POPS the accelerator exhibits an energy efficiency of ~94 fJ/OP.

**References of the Supplementary Materials**